\documentclass{ws-ijmpd}
\usepackage[super,compress]{cite}
\usepackage{hyperref}
\usepackage{mathtools}
\usepackage{float}
\usepackage{booktabs}
\usepackage{graphicx}
\usepackage{tabularx}
\usepackage{orcidlink}

\begin{document}


%
\catchline{}{}{}{}{}
%

\title{Traversable wormholes in the traceless $f(R,T)$ gravity}

\author{Parbati Sahoo\orcidlink{0000-0002-5043-745X}}

\address{Astrophysics and Cosmology Research Unit, \\ School of Mathematics, Statistics and Computer Science,\\ University of KwaZulu-Natal, Private Bag X54001,Durban 4000, South Africa,\\
sahooparbati1990@gmail.com}

\author{P.H.R.S. Moraes\orcidlink{0000-0002-8478-5460}}

\address{Universidade de S\~ao Paulo (USP), Instituto de Astronomia, \\ Geof\'isica e Ci\^encias Atmosf\'ericas (IAG), \\ Rua do Mat\~ao 1226, Cidade Universit\'aria, 05508-090 S\~ao Paulo, SP, Brazil;\\
Instituto Tecnol\'ogico de Aeron\'autica (ITA), Departamento de F\'isica,\\ 12228-900, S\~ao Jos\'e dos Campos, SP, Brazil,\\
moraes.phrs@gmail.com}

\author{Marcelo M. Lapola\orcidlink{0000-0001-5541-4292}}

\address{Instituto Tecnol\'ogico de Aeron\'autica, Departamento de F\'isica,\\ 12228-900, S\~ao Jos\'e dos Campos, SP, Brazil,\\
marcelo.lapola@gmail.com}

\author{P.K. Sahoo\orcidlink{0000-0003-2130-8832}}

\address{Department of Mathematics, Birla Institute of Technology and Science-Pilani, \\ Hyderabad Campus, Hyderabad-500078, India\\
pksahoo@hyderabad.bits-pilani.ac.in}

\maketitle

\begin{history}
\received{02 March 2021}
\end{history}

\begin{abstract}

Wormholes are tunnels connecting different regions in space-time. They were obtained originally as a solution for Einstein's General Theory of Relativity and according to this theory they need to be filled by an exotic kind of anisotropic matter. In the present sense, by ``exotic matter'' we mean matter that does not satisfy the energy conditions. In this article we propose the modelling of traversable wormholes (i.e., wormholes that can be safely crossed) within an alternative gravity theory that proposes an extra material (rather than geometrical) term in its gravitational action, namely the traceless $f(R,T)$ theory of gravitation, with $R$ and $T$ being respectively the Ricci scalar and trace of the energy-momentum tensor. Our solutions are obtained from well-known particular cases of the wormhole metric potentials, namely redshift and shape functions. In possession of the solutions for the wormhole material content, we also apply the energy conditions to them. The features of those are carefully discussed.

\end{abstract}


\tableofcontents

\section{Introduction}\label{sec:int}

Wormholes are outstanding solutions predicted by General Relativity Theory \cite{morris/1988,morris/1988b,visser/1995}. They can be used as a shortcut to interstellar travel and also be converted into a time machine. 

 Wormholes were firstly proposed in the literature as a tool for teaching General Relativity \cite{morris/1988}. Over time, the first wormholes searching proposals started to appear. Gravitational lensing is probably the most optimistic technique to detect wormholes \cite{jusufi/2018,tsukamoto/2016,shaikh/2017,kuhfittig/2014,nandi/2006}, although other possibilities can be appreciated \cite{shaikh/2018,tsukamoto/2012,ohgami/2015,ohgami/2016,li/2014,sabin/2017} , including gravitational waves methods \cite{nandi/2017,aneesh/2018,kirilov/2021}. Despite the efforts, no wormhole detection has been registered so far.

 Morris and Thorne have shown that wormholes can be traversable, that is, safely crossed by travelers. This implies in the aforementioned interstellar travel alternative feature of wormholes. 
 On this regard, the possibility of finding wormholes in the galactic halo regions can be appreciated in References \cite{rahaman/2014,rahaman/2014b,rahaman/2016}.


The absence of horizons (departing from black holes features) is an essential property of traversable wormholes \cite{morris/1988,visser/1995}.  A horizon, if present, would prevent the two-way travel through the wormhole.

According to General Relativity, another important property of traversable wormholes is that they must be filled by matter violating {\bf the null} energy condition \cite{morris/1988b}. 

The fluid permeating wormholes is also expected to be anisotropic, a property that has also been considered in stellar configurations \cite{mak/2003,mak/2002,harko/2002,gleiser/2004,dev/2003}. Of course, we still do not know the wormholes equation of state, and because of that, different proposals can be seen in the literature, as one can check, for instance, Refs.\cite{garattini/2007,sharif/2012,sharif/2019,riazi/2000}. The eventual wormhole detection will certainly constrain these models.

 The aforementioned wormhole exotic matter issue could apparently be evaded when considering these objects in {\it extended theories of gravity} \cite{wands/1994,sinha/2009}. Those extend Einstein's General Relativity, by inserting in the Einstein-Hilbert action some non-linear dependency on the Ricci scalar $R$ or general dependency on other scalars.

Extended theories of gravity are mainly used with the purpose of circumventing some current cosmological problems, such as dark energy \cite{bloomfield/2013,tsujikawa/2010} and dark matter \cite{choudhury/2016}. However,  extended gravity applications to wormhole physics have also recently appeared \cite{moraes/2017,moraes/2019b,elizalde/2018,canfora/2008}. 

As it has been shown in \cite{moraes/2017,moraes/2019b,elizalde/2018}, for instance, particularly the $f(R,T)$ gravity has been a possibility to approach wormholes (it has also been used to treat the dark sector of the universe \cite{alvarenga/2013,shabani/2013,chakraborty/2013,baffou/2015,zaregonbadi/2016}). In the $f(R,T)$ theory of gravity formalism, $f(R,T)$ is a generic function of $R$ and the trace of the energy-momentum tensor $T$, with the latter being inserted in the gravitational action with the purpose of describing imperfect fluids effects. The $T$-dependence may also be related with quantum corrections \cite{harko/2011}. 

Some issues regarding the physical significance of the $T$ dependent terms in $f(R,T)$ gravity were pointed in \cite{fisher/2019} and later corrected in \cite{harko/2020}. In \cite{moraes/2019}, an alternative to the arbitrariness on the choice of the $f(R,T)$ matter lagrangian was proposed. From such a novel formalism, traceless field equations were obtained. This approach was later extended in \cite{carvalho/2021}.

In the present article we will obtain and analyse wormhole solutions in the  aforementioned recently proposed traceless $f(R,T)$ theory of gravitation \cite{moraes/2019}. The article is organized as follows: in Section \ref{sec:frt} we present the traceless $f(R,T)$ gravity. In Section \ref{sec:whmemt} we present the wormhole metric, the conditions that must be satisfied by its potentials as well as the wormhole energy-momentum tensor and the reseulting field equations in the concerned gravity theory. The solutions for the wormhole shape function, matter-energy density and pressures in the traceless $f(R,T)$ gravity are presented in Section \ref{sec:whs} as well as the referred energy conditions. We discuss our results in Section \ref{sec:d}. 

\section{The $f(R,T)$ and traceless $f(R,T)$ theories of gravity}\label{sec:frt}

The $f(R,T)$ gravity theory starts from the action \cite{harko/2011}

\begin{equation}\label{frt1}
	S=\int d^4x\sqrt{-g}\left[\frac{f(R,T)}{16\pi}+\mathcal{L}\right],
\end{equation}
with $g$ being the metric determinant, $f(R,T)$ being a function of the argument and $\mathcal{L}$ the matter Lagrangian density.

By assuming $f(R,T)=R+\mathcal{F}(T)$, which means to particularize the $R$ dependence of the theory to be the same as in General Relativity while letting the field equations to depend generically on $T$, through the general function $\mathcal{F}$, the variational principle applied in \eqref{frt1} yields 

\begin{equation}\label{frt2}
	G_{\mu\nu}=8\pi T_{\mu\nu}+\frac{\mathcal{F}g_{\mu\nu}}{2}+\frac{d\mathcal{F}}{dT}(T_{\mu\nu}-\mathcal{L}g_{\mu\nu}).
\end{equation}
In \eqref{frt2}, $G_{\mu\nu}$ is the Einstein tensor, $T_{\mu\nu}$ is the energy-momentum tensor and $g_{\mu\nu}$ is the metric.

It is clear, then, that for this case, any divergence from General Relativity appears on the {\it rhs} of the field equations (matter-energy) rather than on their {\it lhs} (curvature). In this case it is worth to briefly mention the physical interpretation of the extra terms on \eqref{frt2}. The $T$-dependence of the theory can be interpreted as due to at least one of the following factors: fluid imperfections, extra fluid, quantum effects, effective cosmological constant (check \cite{moraes/2019} and references therein). 

In \cite{moraes/2019}, the explicit dependence of the $f(R,T)$ gravity field equations on the choice of the matter Lagrangian density was profoundly discussed. The fact that different choices for $\mathcal{L}$ yield different field equations is unsatisfactory in view of a unique mathematical description of natural phenomena. Still in \cite{moraes/2019}, a method to evade this problem was proposed from the isolation of $\mathcal{L}$ in the trace equation of \eqref{frt2}, namely

\begin{equation}\label{frt3}
	-R=2(4\pi T+\mathcal{F})+\frac{d\mathcal{F}}{dT}(T-4\mathcal{L}),
\end{equation} 
and consequent substitution of it in \eqref{frt2}, which yields the following field equations:

\begin{equation}\label{frt4}
	R_{\mu\nu}-\frac{Rg_{\mu\nu}}{4}=\left(8\pi+\frac{d\mathcal{F}}{dT}\right)T_{\mu\nu}-\left(2\pi T+\frac{1}{4}\frac{d\mathcal{F}}{d\ln T}\right)g_{\mu\nu},
\end{equation}
with $R_{\mu\nu}$ being the Ricci tensor. 

The field equations \eqref{frt4} not only evade the need for choosing a particular matter Lagrangian density but also attain a unimodular theory of gravity \cite{ng/1991,tiwari/1993,alvarez/2005}. Note that by simply taking $\mathcal{F}=0$ one recovers the original field equations of unimodular gravity proposed by Einstein \cite{einstein/1919}. In this way, the theory in \eqref{frt4} ``corrects'' the unimodular gravity by inserting extra material terms in it, and those are likewise motivated by the aforementioned factors.  Remarkably, Eq.\eqref{frt4} keeps traceless for any choice of the function $\mathcal{F}$. 

 It is worth noting that in principle the energy-momentum tensor does not conserve in Eq.\eqref{frt4}. The extra terms in \eqref{frt4} can be related to an extra force that changes the geodesic equation and forbid the energy-momentum tensor to conserve in a first analysis \cite{harko/2011,bertolami/2007,harko/2010,harko/2014}. In a cosmological aspect, this could be related to creation of matter throughout the universe evolution \cite{harko/2014b}. Anyhow, in order to keep the energy-momentum tensor conserved in Eq.\eqref{frt4} one can follow the approach presented in \cite{carvalho/2020,dos_santos_jr/2019}.

As it has been mentioned in \cite{moraes/2019}, the theory described in Eq.\eqref{frt4}, which we are referring to as traceless $f(R,T)$ gravity, raises as a new possibility to investigate the universe in its different regimes, such as stellar and galactic astrophysics and cosmology. Here we choose to apply the traceless $f(R,T)$ gravity to wormholes.

\section{Wormhole field equations}\label{sec:whmemt}

In this section we are going to model wormholes in the traceless $f(R,T)$ gravity theory. In order to do so we are going to present the wormhole metric and energy-momentum tensor and substitute them in the traceless $f(R,T)$ gravity field equations for a particular choice of the function $\mathcal{F}$. 

The static spherically symmetric wormhole metric reads \cite{morris/1988,morris/1988b,visser/1995}

\begin{equation}\label{m1}
ds^2=e^{a(r)}dt^2-\frac{dr^2}{1-\frac{b(r)}{r}}-r^2(d\theta^2+sin^2\theta d\phi^2),
\end{equation}
where $a(r)$ and $b(r)$ are the redshift function and shape function, respectively. Moreover, $r$ is the radial coordinate, which increases from a minimum radius value to $\infty$, i.e. $r_0\leq r <\infty$, where $r_0$ is known as the throat radius. A flaring out condition of the throat is considered as an important condition to have a typical wormhole solution, such that $\frac{b-b'r}{b^2}>0$ \cite{morris/1988}, with primes indicating radial derivatives, and at the throat, $r=r_0=b(r_0)$. Also, in order to have wormhole solutions, the condition $b'(r_0)<1$ must be satisfied. The absence of horizons and singularities is ensured when the redshift function $a(r)$ is finite and nonzero throughout the space-time \cite{morris/1988}.

The energy-momentum tensor of wormholes reads
\begin{equation}\label{m2}
T_{\mu\nu}=(\rho+p_t)u_\mu u_\nu-p_t g_{\mu\nu}+ (p_r-p_t)X_\mu X_\nu,
\end{equation}
with $\mu,\nu=0,1,2,3$ and where $\rho$, $p_r$, and $p_t$ are the energy density, radial pressure and tangential pressure, respectively, $u_\mu$ and $X_\mu$ are the four-velocity vector and radial unit four-vector, satisfying the relations \textbf{$u_\mu u^\mu = 1$ and $X_\mu X^\mu = -1$}. The trace of $T_{\mu\nu}$ is $T=\rho-p_r-2p_t$. 

For $\mathcal{F}=2\lambda T$, with constant $\lambda$, the traceless $f(R,T)$ gravity field equations read

\begin{equation}\label{e4}
R_{\mu\nu}-\frac{Rg_{\mu\nu}}{4}=(8\pi +2\lambda )T_{\mu\nu}-\biggl(\frac{4\pi+\lambda}{2}\biggr)Tg_{\mu\nu}.
\end{equation}

Equation (\ref{e4}) for the wormhole metric (\ref{m1}) and energy-momentum tensor \eqref{m2}, with the choice of constant redshift function, reads

\begin{eqnarray}
\frac{b'(r)}{r^2}= (12\pi+3\lambda)\rho+ (4\pi+\lambda)(p_r+2p_t),\label{fe1}\\
\frac{2b(r)-rb'(r)}{r^3}= -(12\pi+3\lambda)p_r-(4\pi+\lambda)(\rho-2p_t),\label{fe2}\\
\frac{-b(r)}{r^3}= -2(4\pi+\lambda)p_t-(4\pi+\lambda)(\rho-p_r).\label{fe3}
\end{eqnarray}
 
\section{Wormhole solutions and energy conditions}\label{sec:whs}

Wormholes in General Relativity lead to a violation of causality with the mathematical prediction of the occurrence of exotic matter inside them, i.e., matter that violates the energy conditions, which will be described below. 

The energy conditions are a variety of ways of making the notion of locally positive energy density more precise. They are described as the following  \cite{morris/1988,morris/1988b,visser/1995}:

$\bullet$ The strong energy condition (SEC) says that gravity should be always attractive, or in terms of energy-momentum tensor it reads $\rho+ \Sigma_j p_j\geq0$, $\forall j$ .  

$\bullet$ The dominant energy condition (DEC) is an indication that the energy density measured by any observer should be non-negative, which leads to $\rho\geq|p_j|$, $\forall j$. 

$\bullet$ The weak energy condition (WEC) shows that the energy density measured by any observer should be always non-negative, i.e., $\rho\geq0$ and $\rho + p_j\geq0$, $\forall j$. 

$\bullet$ The null energy condition (NEC) is a minimum requirement from SEC and WEC, i.e. $\rho+p_j\geq0$, $\forall j$. 
The violation of NEC implies that all the above energy
conditions are not validated. 

In our specific case for wormholes, $p_j$ can be the radial and tangential pressures, $p_r$ and $p_t$. So, in the following, considering an ansatz for the shape function $b(r)$, we will investigate the WEC mentioned above. After that, we will check the others with a help of an equation of state, as follows.

\subsection{Particular choice of shape function}

In the literature, different shape functions are employed to study wormhole geometry and properties. For example: logarithmic shape function \cite{Godani}, exponential shape function \cite{SahooKulkarni}, power law \cite{Lobo/2009} and hybrid (combination of power law and exponential) shape function \cite{sahoo}. Motivated by these forms, we consider here the following hybrid shape function in the form \cite{sahoo}
\begin{equation}\label{b1}
b(r)=r_0^m e^{\alpha(r_0-r)}r^{1-m},
\end{equation}
 where $m$ and $\alpha$ are constants. The behaviour of $b(r)$ and its radial derivative with $m=2$ and $\alpha=1$ can be seen in Fig.1 below. 

\begin{figure}[h]
\centering
\includegraphics[scale=0.45]{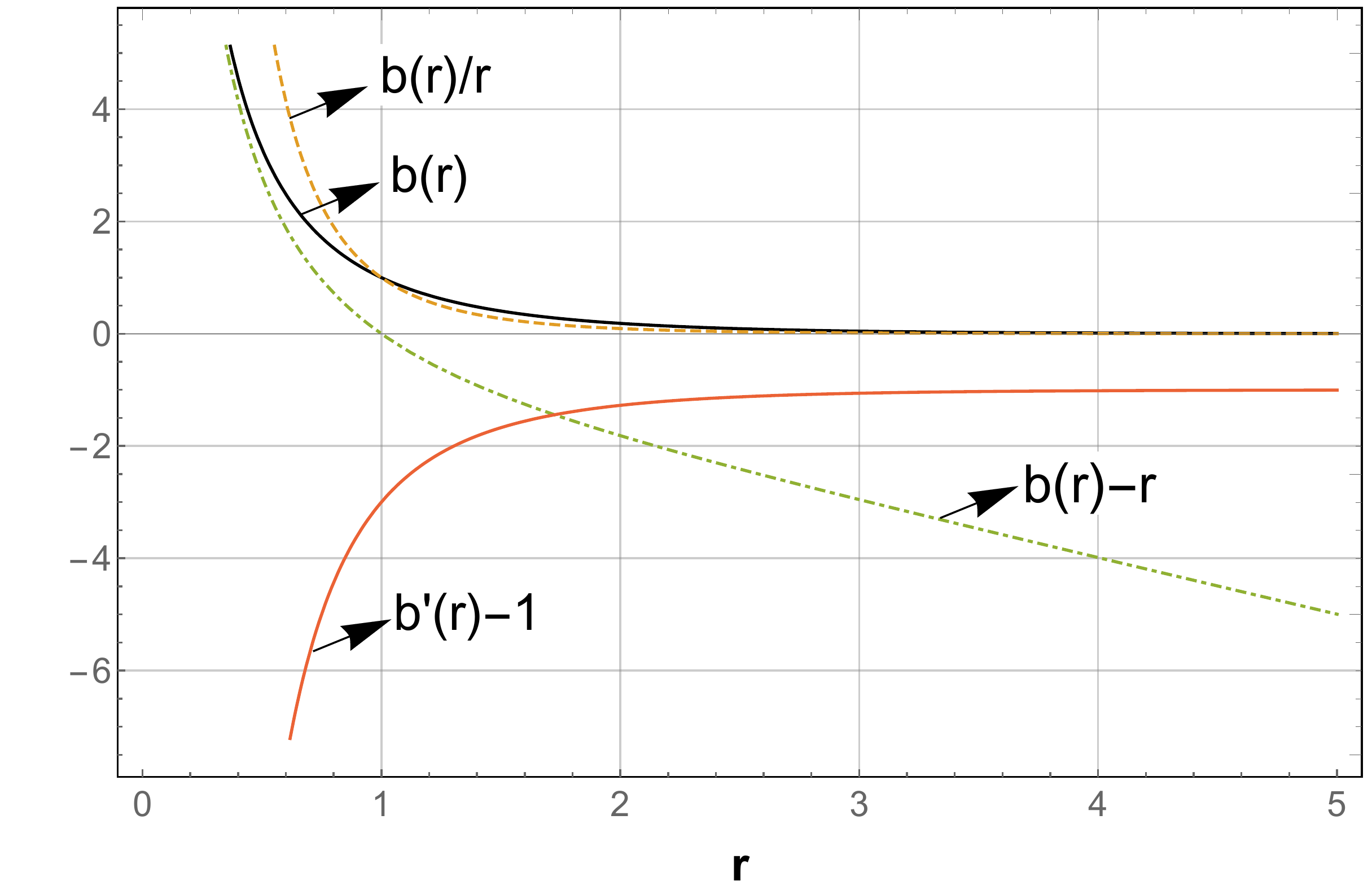}
\caption{Behavior of the shape function and its radial derivative with $r_0=1$, $\alpha=1$, and $m=2$.}
\label{fig1}
\end{figure}

Equations (\ref{fe1})-(\ref{fe3}) with the choice of (\ref{b1}) can be rewritten as

\begin{equation}\label{ee3}
\rho+p_r=-\frac{e^{1-r} (r+2)}{2 (\lambda +4 \pi ) r^4},
\end{equation}

\begin{equation}\label{ee4}
\rho+p_t=-\frac{e^{1-r}}{4 (\lambda +4 \pi ) r^3}.
\end{equation}

To get the positive behavior in (\ref{ee3}) and (\ref{ee4}), the $\lambda$ value must be restricted to $\lambda <-4 \pi$. The graphical representation of their positive behavior is given in the Figures 2 and 3 below.

\begin{figure}[h]
\centering
\begin{minipage}{100mm}
\includegraphics[width=90 mm]{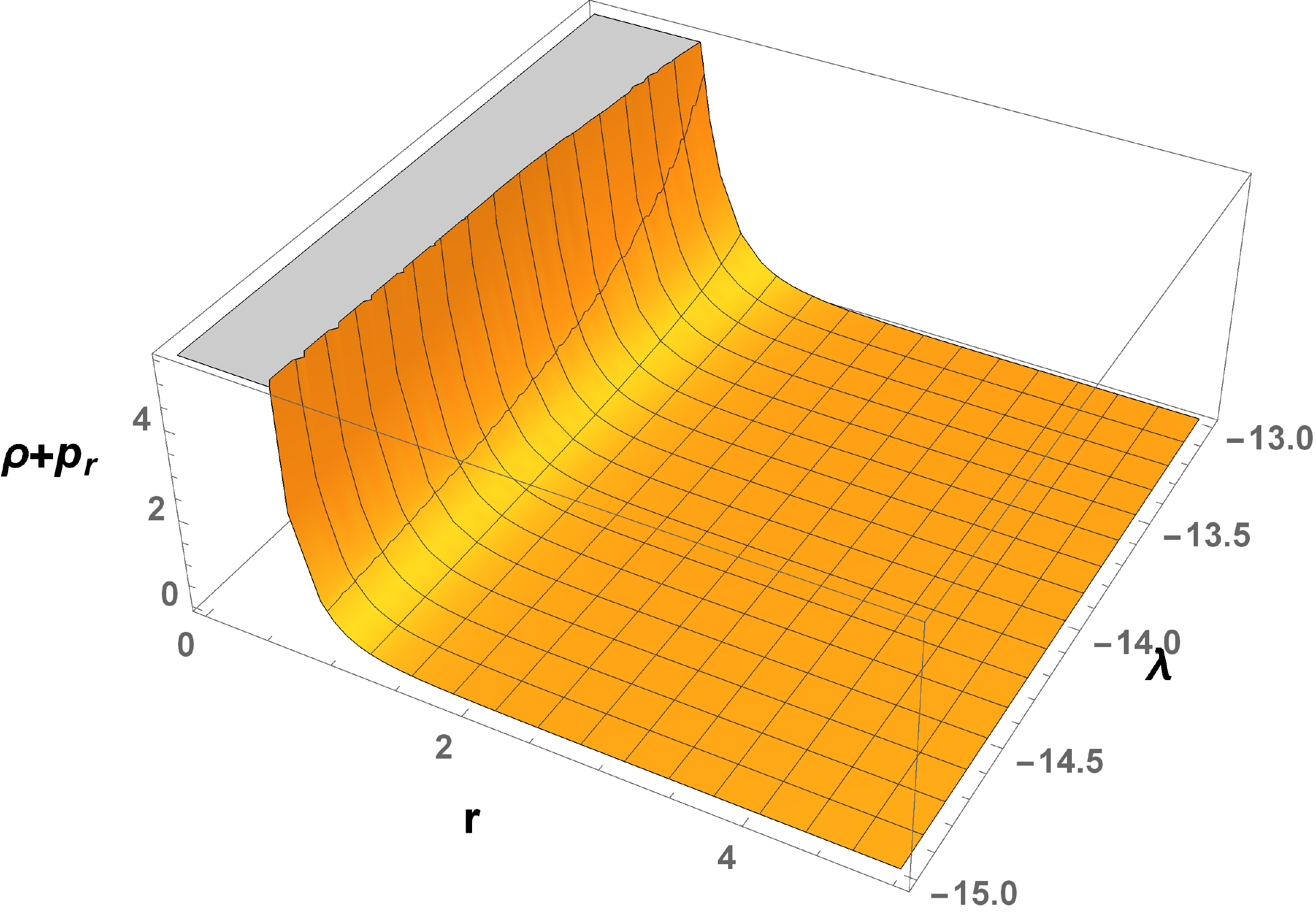}
\caption{Behavior of $\rho+p_r$ with $r_0=1$, $\lambda<-4\pi$, $\alpha=1$, and $m=2$.}
\label{fig1}
\end{minipage}
\hfill
\begin{minipage}{100mm}
\includegraphics[width=90mm]{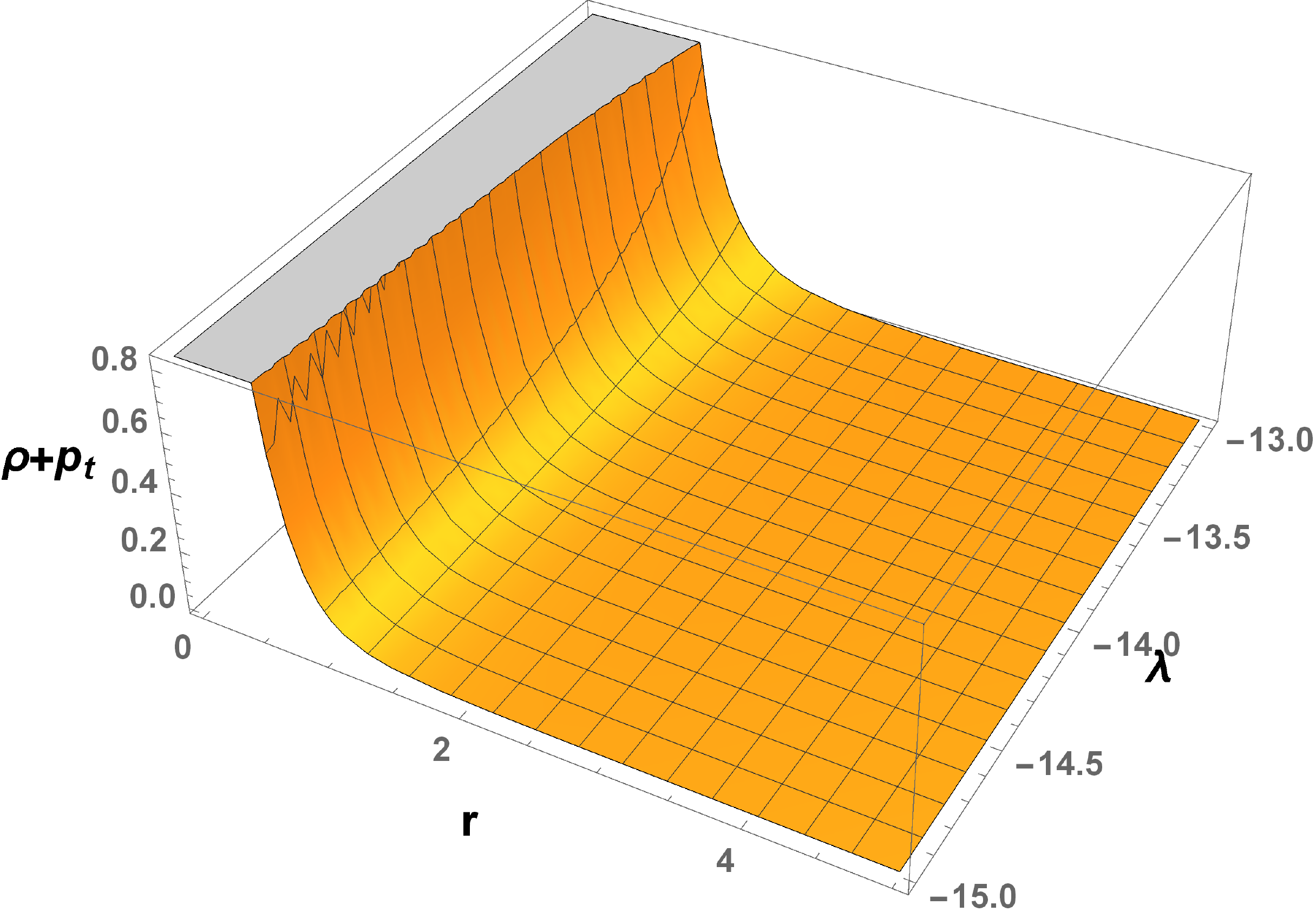}
\caption{Behavior of $\rho+p_t$ with $r_0=1$, $\lambda<-4\pi$, $\alpha=1$, and $m=2$.}
\label{fig2}
\end{minipage}
\end{figure} 
  
\subsection{Particular choice of equation of state}

Let us concentrate on Equations \eqref{ee3}-\eqref{ee4}, which are an implicit relation between matter components obtained from the field equations \eqref{fe1}-\eqref{fe3} with four unknown and two equations. After considering the specific choice of shape function, we have two equations with three left unknowns, for which we have to take one more arbitrary choice in order to obtain the explicitly detailed behavior of matter components in the wormhole constructions. Let us keep the above choice of shape function and consider the equation of state

\begin{equation}
p_r=\omega\rho,
\end{equation}  
with $\omega$ as a constant. 

Now we obtain

\begin{equation}\label{ee5}
\rho=-\frac{b(r)-rb'(r)}{2 (\lambda +4 \pi ) r^3 (\omega +1)}=-\frac{e^{1-r} (r+2)}{2 (\lambda +4 \pi ) r^4 (\omega +1)},
\end{equation}
\begin{equation}\label{ee6}
p_r= -\frac{\omega[b(r)-rb'(r)]}{2 (\lambda +4 \pi ) r^3 (\omega +1)}=-\frac{e^{1-r} (r+2)\omega}{2 (\lambda +4 \pi ) r^4 (\omega +1)},
\end{equation}

\begin{equation}\label{ee7}
p_t= \frac{b(r) [\omega +3]+rb'(r) (\omega -1)}{4 (\lambda +4 \pi ) r^3 (\omega +1)}=\frac{e^{1-r} [4+r (1-\omega)]}{4 (\lambda +4 \pi ) r^4 (\omega +1)}.
\end{equation}

In Equation (\ref{ee5}), in order to achieve a positive density, the restrictions on $\lambda$ and $\omega$ are given as $\lambda >-4\pi$ and $\omega<-1 $ or $\lambda < -4\pi$, $\omega > -1$.    

Combining Eqs. (15)-(17), the energy conditions are obtained as below 
\begin{eqnarray}
\rho+p_r=-\frac{e^{1-r} (r+2)}{2 (\lambda +4 \pi ) r^4},\\
\rho+p_t=-\frac{e^{1-r}}{4 (\lambda +4 \pi ) r^3},\\
\rho-p_r=\frac{e^{1-r} (r+2) (\omega -1)}{2 (\lambda +4 \pi ) r^4 (\omega +1)},\\
\rho-p_t=\frac{e^{1-r} (r (\omega -3)-8)}{4 (\lambda +4 \pi ) r^4 (\omega +1)},\\
\rho+p_r+2p_t=-\frac{e^{1-r} (r \omega +\omega -1)}{(\lambda +4 \pi ) r^4 (\omega +1)}.
\end{eqnarray}

For $\lambda < -4\pi$, $\omega > -1$, the graphical representation of all the energy conditions is given in Figures \ref{figa}-\ref{figg} below. Note that the horizontal axis for each of the plots is considered for the radial coordinate and the physical units are \textit{km}.

\begin{figure}[h]
\centering
\begin{minipage}{100mm}
\includegraphics[width=90 mm]{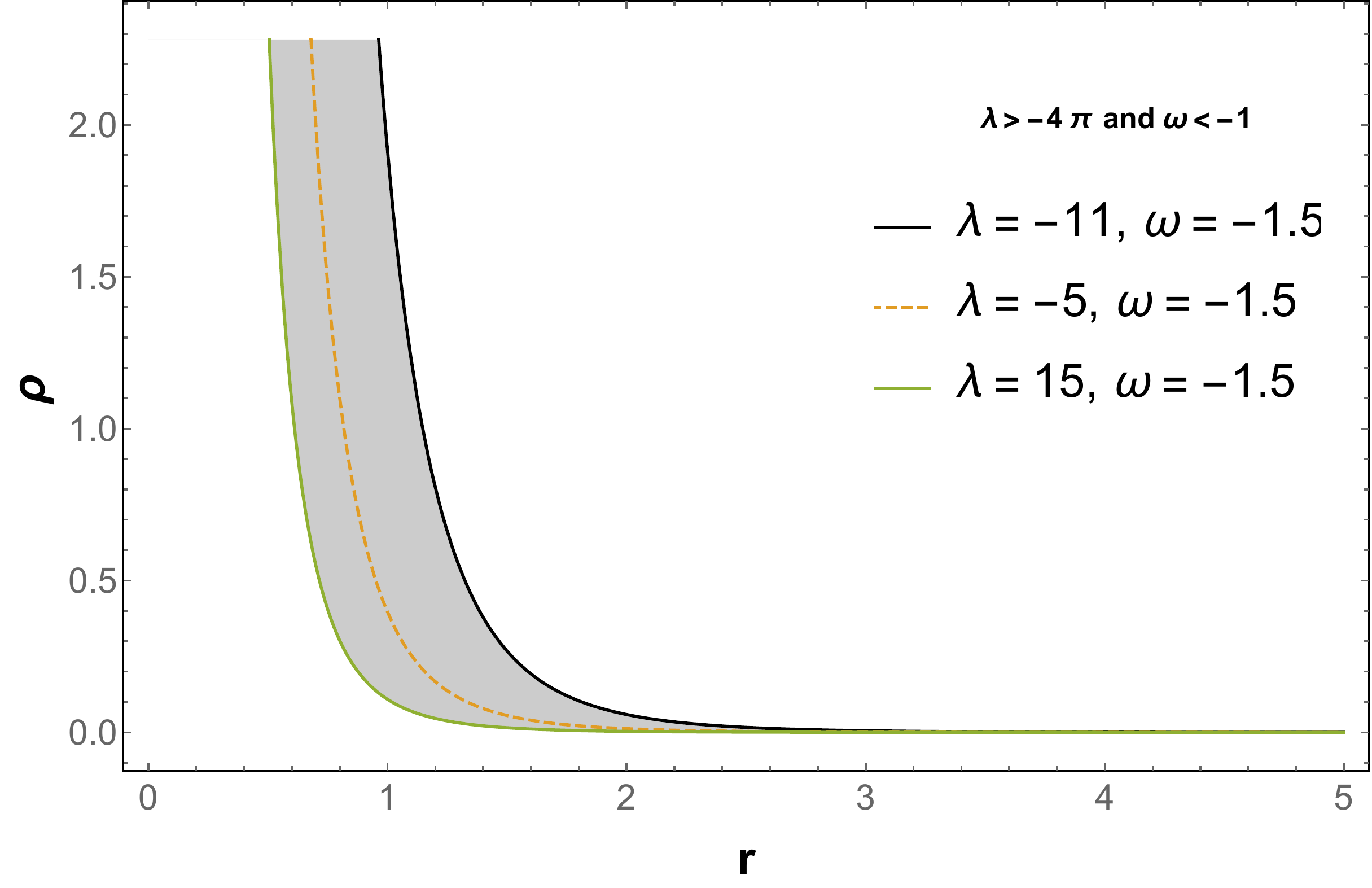}
\caption{Behavior of $\rho$ with $r_0=1$, $\lambda >-4\pi$, $\omega<-1$, $\alpha=1$, and $m=2$.} \label{figa}
\end{minipage}
\hfill
\begin{minipage}{100mm}
\includegraphics[width=90 mm]{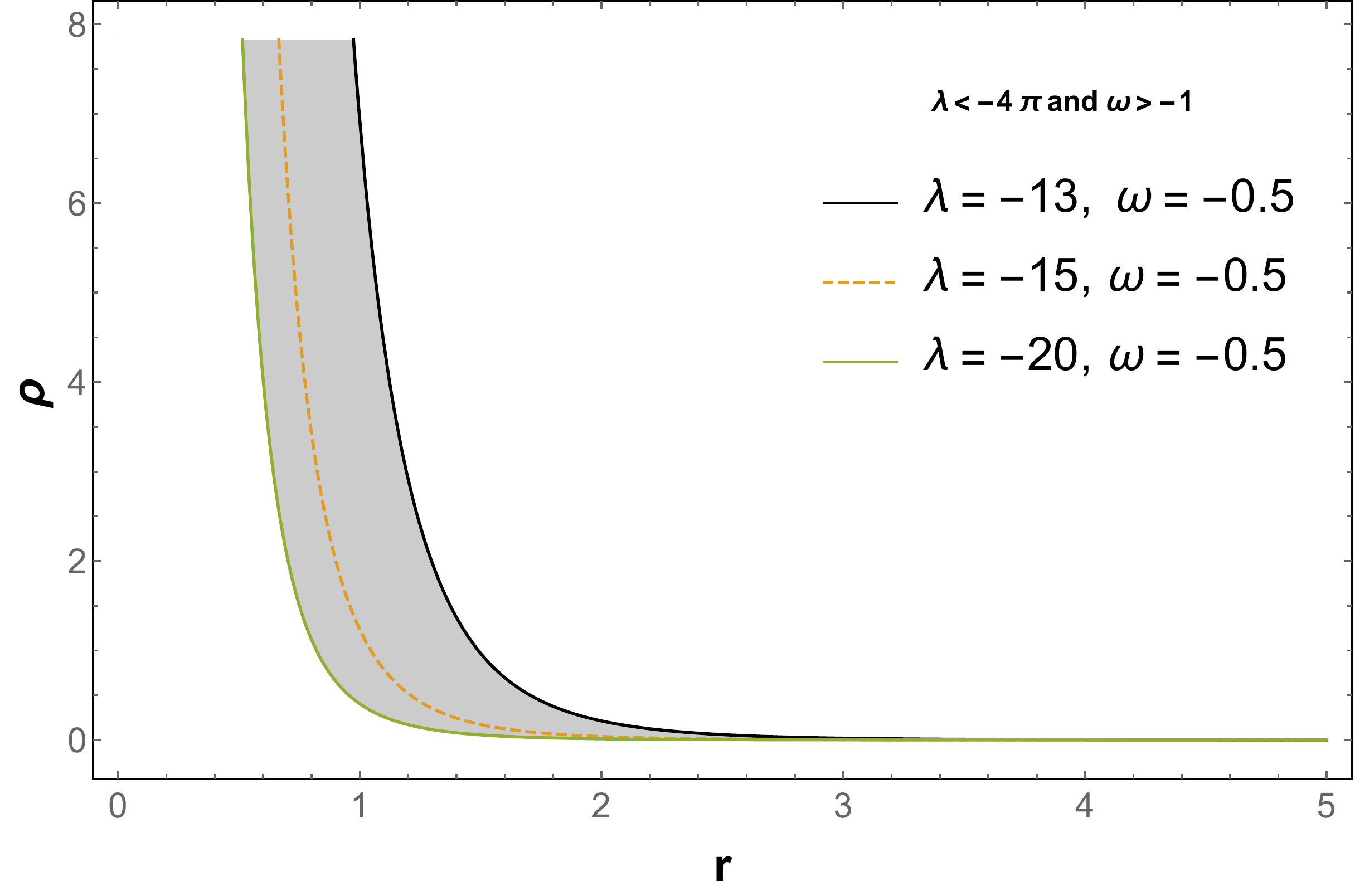}
\caption{Behavior of $\rho$ with $r_0=1$, $\lambda < -4\pi$, $\omega > -1$, $\alpha=1$, and $m=2$.} \label{figb}
\end{minipage}
\end{figure}

\begin{figure}[H]
\centering
\includegraphics[width=90 mm]{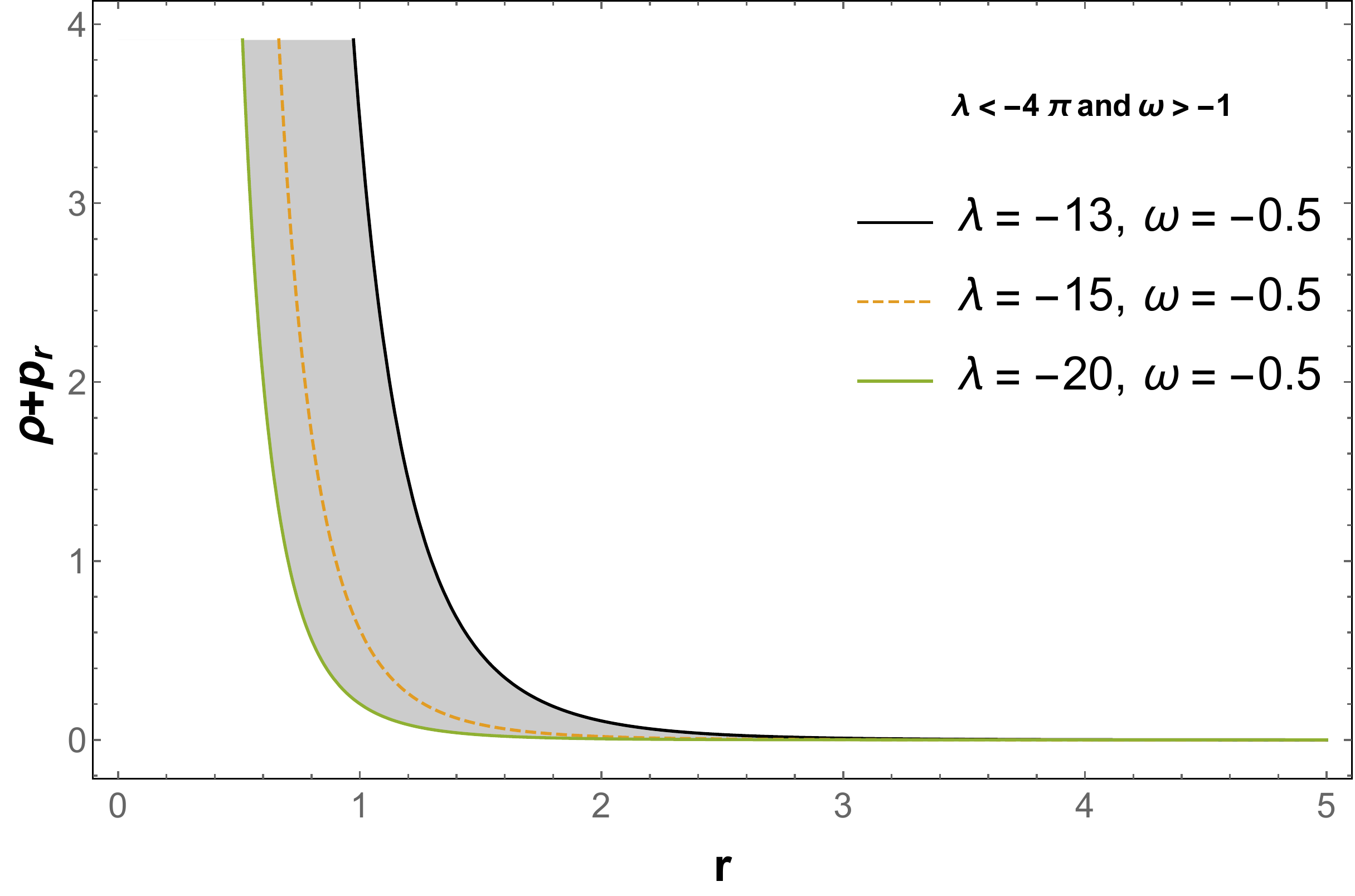}
\caption{Behavior of $\rho+p_r$ with $r_0=1$, $\alpha=1$, and $m=2$.} \label{figc}
\end{figure}

\begin{figure}[H]
\centering
\begin{minipage}{100mm}
\includegraphics[width=90 mm]{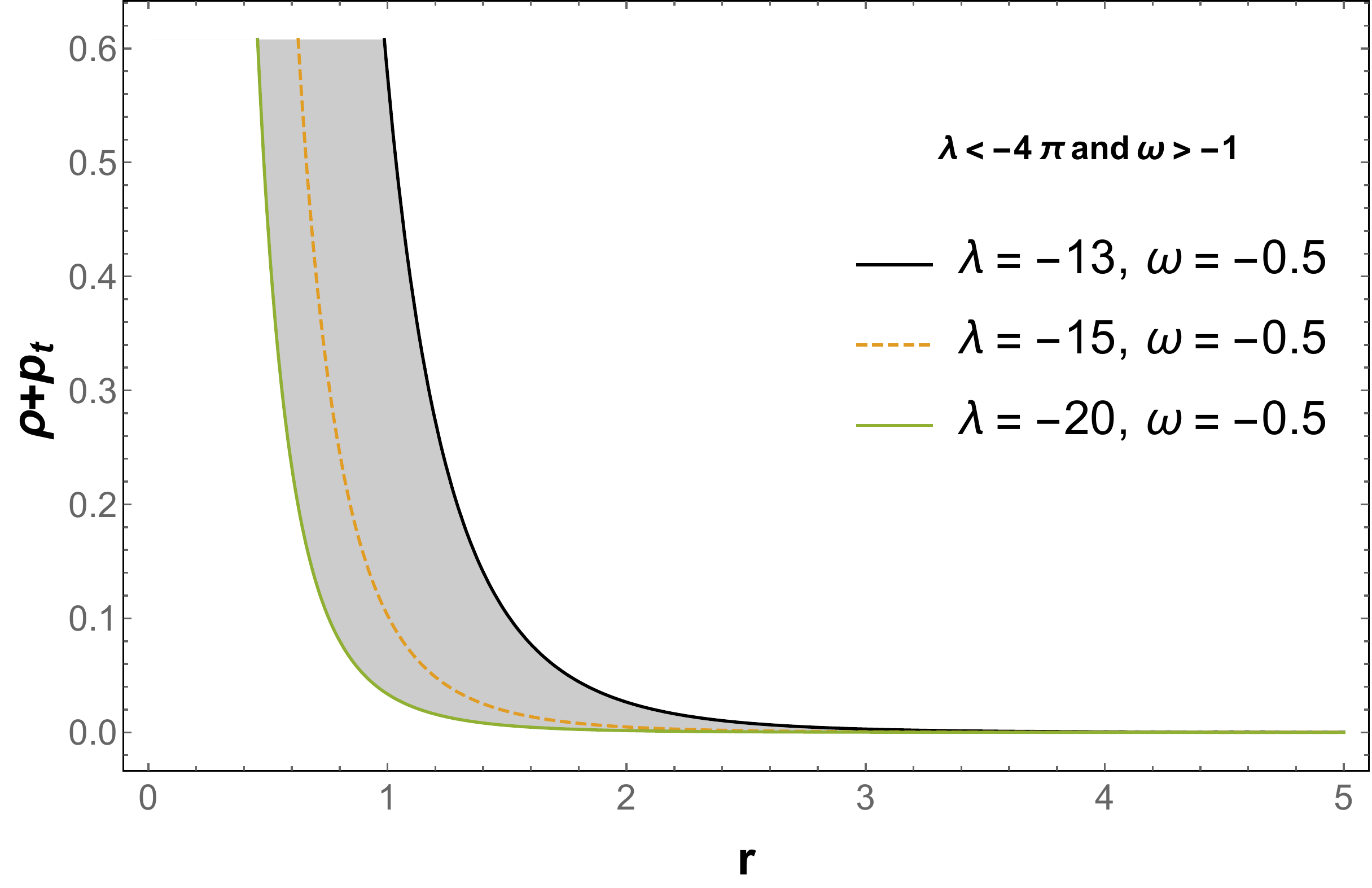}
\caption{Behavior of $\rho+p_t$ with $r_0=1$, $\alpha=1$, and $m=2$. } \label{figd}
\end{minipage}
\hfill
\begin{minipage}{100mm}
\includegraphics[width=90 mm]{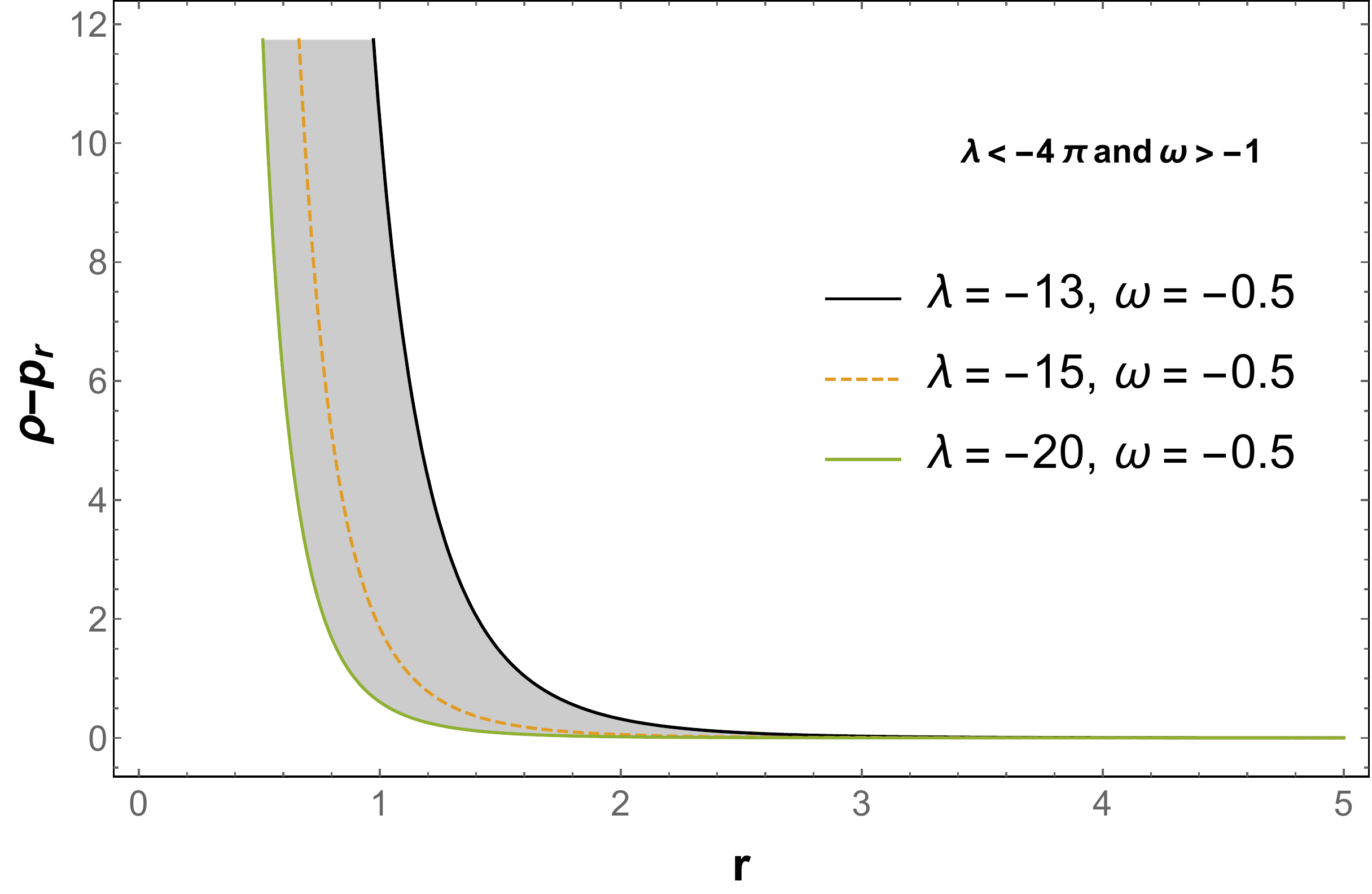}
\caption{Behavior of $\rho-p_r$ with $r_0=1$, $\alpha=1$, and $m=2$.} \label{fige}
\end{minipage}
\end{figure}

\begin{figure}[H]
\centering
\includegraphics[width=90 mm]{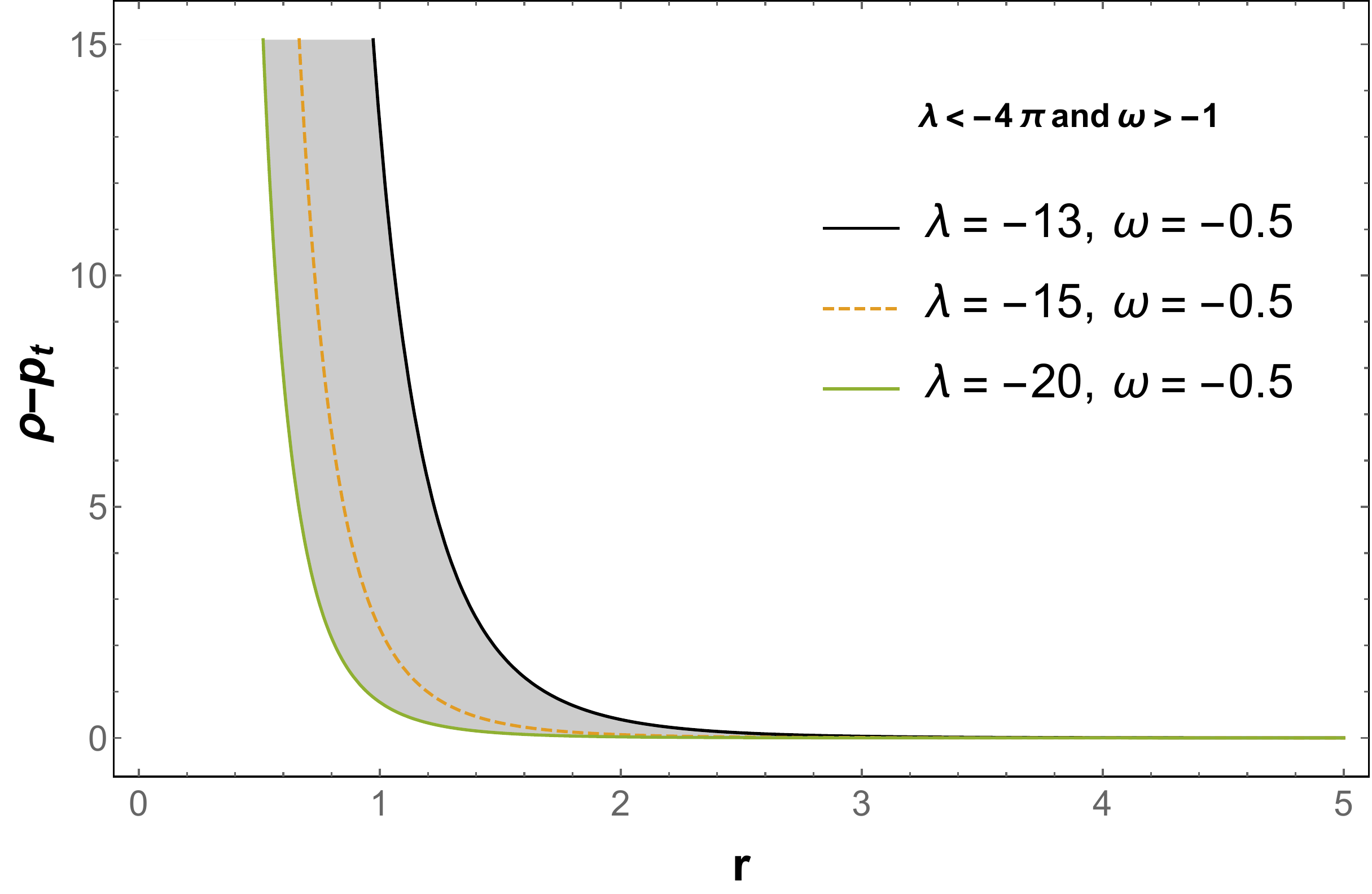}
\caption{Behavior of $\rho-p_t$ with $r_0=1$, $\alpha=1$, and $m=2$.} \label{figf}
\end{figure}

\begin{figure}[H]
\centering
\includegraphics[width=90 mm]{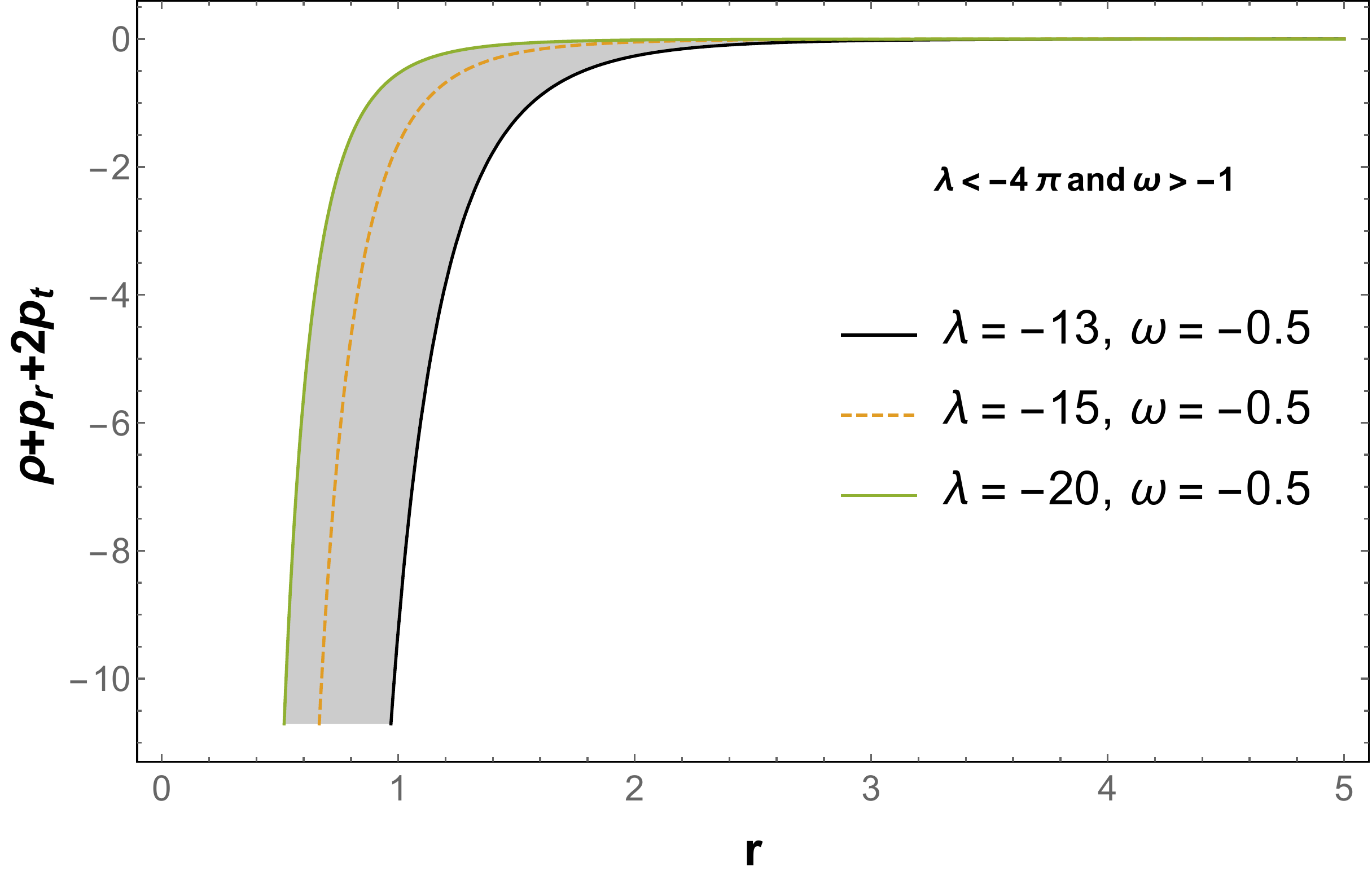}
\caption{Behavior of $\rho+p_r+2p_t$ with $r_0=1$, $\alpha=1$, and $m=2$.} \label{figg}
\end{figure}
The behavior of all energy conditions with respect to the all given parameter restrictions is summarized in the table below.
\begin{table}[H]
\begin{center}
\begin{tabular}{ |c|c|c| } 
 \hline
  & $\lambda >-4\pi$, $\omega<-1 $ & $\lambda < -4\pi$, $\omega > -1$ \\ 
  \hline
 $\rho+p_r$ & $<0$ & $>0$ \\ 
\hline
 $\rho+p_t$ & $<0$ & $>0$ \\ 
 \hline 
 $\rho-p_r$ & $>0$ & $>0$ \\
 \hline 
 $\rho-p_t$ & $>0$ & $>0$ \\
 \hline 
 $\rho+p_r+2p_t$ & $<0$ & $<0$ \\ 
 \hline
\end{tabular}
\caption {Behaviour of energy conditions for $\lambda >-4\pi$, $\omega<-1 $ and $\lambda < -4\pi$, $\omega > -1$.}
\end{center}
\label{tb1}
\end{table}

To quantify the exotic matter present in the throat, we utilize the ``volume integral quantifier" \cite{Visser/0003,Nandi/2004}. For a simple case of spherical symmetry and NEC-violating matter related only to radial component, it is defined as
\begin{equation}\label{e19}
I_V= \int_{r_0}^{\infty} \int_{0}^{\pi} \int_{0}^{2\pi}(\rho+p_r)\sqrt{-g_4}dr d\theta d\phi,
\end{equation} 
or we can rewrite it as  $$I_V=\oint (\rho+p_r)dV= 2\int_{r_0}^{\infty} (\rho+p_r)4\pi r^2 dr.$$
From Equation (18), we obtain
\begin{equation}\label{e20}
I_V=\biggl\{\frac{4 e \pi  \left[r \text{Ei}(-r)+2 e^{-r}\right]}{(\lambda +4 \pi ) r}\biggr\}_{r_0}^{\infty},
\end{equation}
where $Ei(-r)$ is an exponential integral function of $-r$.

\begin{figure}[h]
\includegraphics[scale=0.45]{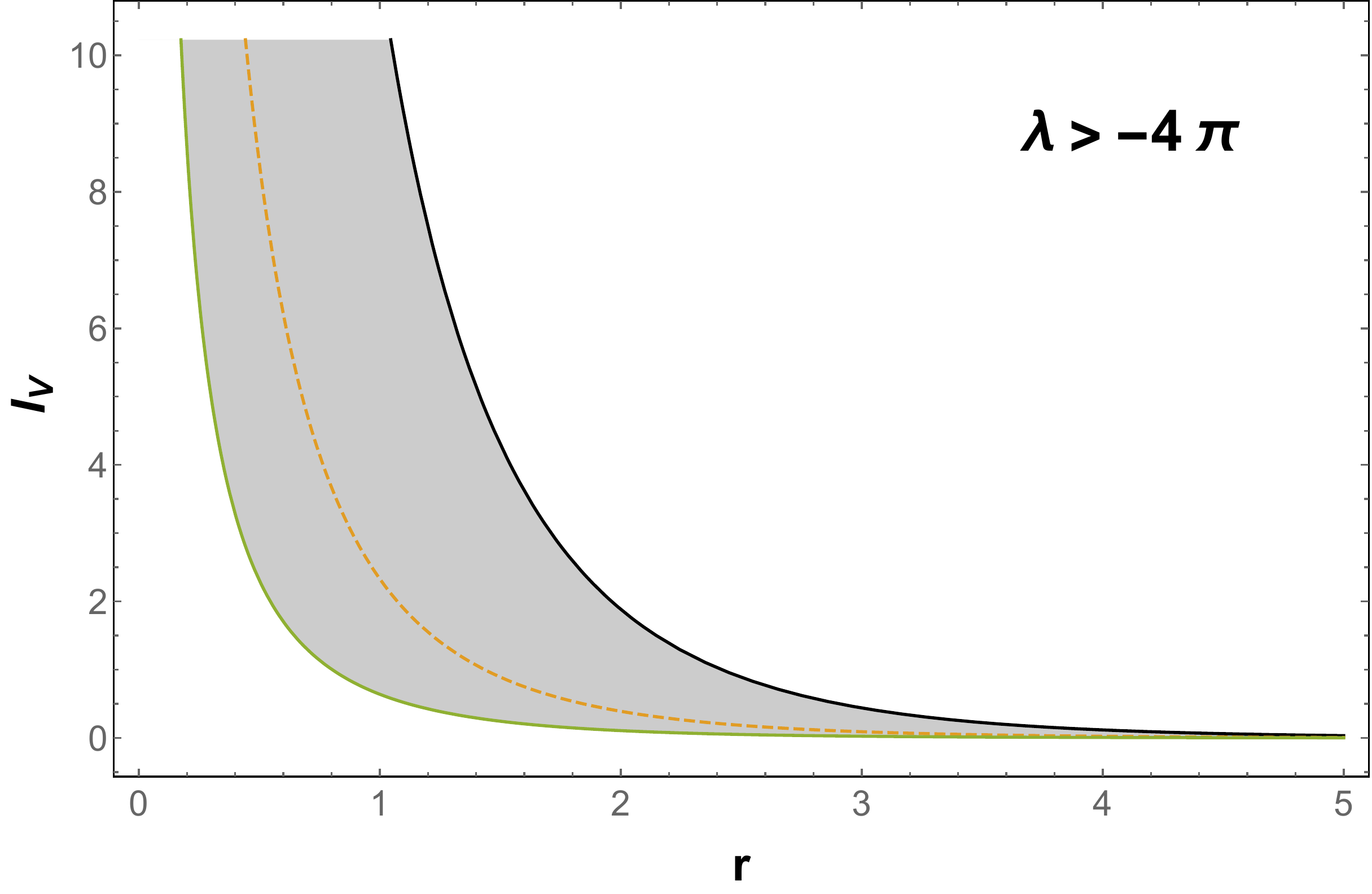}
\caption{Volume integral quantifier ($I_V$) with $\lambda > -4\pi$. }\label{figh}
\end{figure}

\section{Discussion}\label{sec:d}

A spherically symmetric static traversable wormhole was constructed in the present article. The novelty is that we used a traceless version of the $f(R,T)$ theory of gravitation and this combination leads to a satisfactory compliance with the energy conditions. We made this by using a general knowing form for the functional of the function $f(R,T)$, which contains a first order trace of the energy-momentum tensor, namely $f(R,T)=R+2\lambda T$, with $\lambda$ as a constant. The traceless $f(R,T)$ gravity resembles Einstein's unimodular gravity \cite{einstein/1919}, but with corrections in the material sector. 

Initially proposed by Einstein as a traceless portion of General Relativity field equations, in the unimodular gravity, the vacuum energy does not gravitate, so that there is no implication of a unique value for the cosmological constant. That is why unimodular gravity was proposed as a way to solve the cosmological constant problem \cite{weinberg/1989}. Unimodular gravity was recently used to approach the accelerated expansion of the universe, in good agreement with observations  \cite{garcia_aspeitia/2019}. Compact astrophysical objects were investigated in unimodular gravity in \cite{Astorga-Moreno et al./2019}.  On the other hand, the recently proposed traceless $f(R,T)$ gravity still lacks applications. 


 Our traceless $f(R,T)$ gravity wormhole solutions have led us to plot the WEC in  Figures 2 and 3.  From these figures it is clear that the WEC is properly respected within the traceless $f(R,T)$ gravity formalism. In other words, the extra degrees of freedom of traceless $f(R,T)$ gravity, related to $T$-dependent terms, make possible to obey the WEC.

As a brief semantic comment, it is valid to stress that Morris and Thorne in their original article \cite{morris/1988} considered {\it exotic matter} as matter violating the WEC. It is common to see other definitions in the literature, such as the one taken by Visser \cite{visser/1995}, who considers exotic matter as matter violating the NEC.

Figs.4-9 show the obedience of DEC and NEC while Fig.10 depicts the violation of SEC. It is important to state that the cosmological constant, responsible for accelerating the universe expansion in standard cosmology, also violates SEC \cite{visser/1995}. In fact we could say that the present epoch of the universe expansion violates SEC, as SEC says that gravity should always be attractive and the accelerating expansion is a sort of anti-gravitational effect.


As shown in Fig.10, from the analysis of the Eq.(22), the strong energy condition is not satisfied in any of the cases, and the WEC is not obeyed in case with $\lambda >-4\pi$, $\omega<-1 $, which may indicate the presence of some exotic matter in the wormhole throat, even in tiny quantities, as explicitly demonstrated by the calculation of VIQ, and its asymptotic behavior (i.e. $I_V \rightarrow{0}$ as $r \rightarrow{\infty}$) is plotted in Fig.11. 

 It is important to stress here that a more profound discussion about energy conditions in alternative gravity was given in \cite{capozziello/2014,capozziello/2015}, in which the effective energy-momentum tensor of the theory is confronted with the energy conditions, rather than the usual one. In an extended gravity theory with extra material rather than geometrical terms, as the one presented here, in which the extra terms of the effective energy-momentum tensor also depend on $\rho$ and $p$, this approach is rather complicated to be attained, but we shall report it soon in the literature.

In general, an arbitrarily small quantity of exotic matter at wormhole throat will be sufficient to make wormhole traversable.  The quantifier measurement of amount of energy condition violated exotic matter will be consistent with this principle \cite{nandi/2004}.  However, from the calculation of VIQ (\ref{e19}) one sees that $	I_V \rightarrow 0$ when $r \rightarrow \infty$, which makes it arbitrarily small. At the same time one can observe from the graphical representation of $I_V$ that it approaches zero negatively in the phase of $\lambda <-4\pi$ and positively in the phase of $\lambda > -4 \pi$. On the other hand, violation of NEC as well as SEC occurs in the case of $\lambda>-4\pi$ and in the phantom region $\omega<-1$, which may be an indication of presence of exotic matter at the throat of the  wormhole supported by phantom fluid \cite{Sahoo191}. This behavior can be considered as a reference in future to concentrate on the existence of traversable wormhole in traceless $f(R,T)$ gravity.  

\bigskip

{\bf Acknowledgements:}\
PHRSM would like to thank FAPESP grant 2015/08476-0 and CAPES, for financial support. M.M.Lapola thanks CAPES, grant 88882.446980/2019-01, for financial support. PKS acknowledges DST, New Delhi, India for providing facilities through DST-FIST lab, Department of Mathematics, BITS-Pilani, Hyderabad Campus where a part of this work was done. We are very much grateful to the honorable referees and the editor for the illuminating suggestions that 
have significantly improved our work in terms of research quality and presentation.

\end{document}